\documentclass[sigconf,authorversion,nonacm]{acmart}

\def\authnotes{1}

\usepackage{tikz}
\usepackage{stmaryrd}
\usepackage[tworuled,figure,vlined,linesnumbered]{algorithm2e}

\usepackage{latexsym,amsmath,amsfonts,enumerate,alltt} % note for acmart papers: we should not add amssymb as it's already included

\usepackage{caption}
\usepackage{graphicx}
\usepackage{url}
\usepackage{mathtools}

\usepackage{listings}
\usepackage{xcolor}
\usepackage{xspace}

\usepackage{multirow}
\usepackage{tabu}

\usepackage{xspace}

\usepackage{pifont}

\usepackage{enumitem} %for noitemsep,nolistsep
\usepackage[vlined,linesnumbered]{algorithm2e} %from Randy
\usepackage[font=small]{caption}
\usepackage{subcaption}
\usepackage{multirow}

\usepackage{tikz}
\usepackage{pgfplots}
\usetikzlibrary{arrows}
\usetikzlibrary{patterns}
\usepackage{comment}

\usepackage[capitalise]{cleveref}

\usepackage{booktabs}

\newcommand{\mycomment}[1]{}

\newcommand{\ASK}[1] {}

\newcommand{\TBD}[1] {}

\newcommand{\xsrules}[1]{\{\rho_1,...,\rho_k\}}
\newcommand{\calD}{{\mathcal D}}
\newcommand{\calE}{{\mathcal E}}

\newcommand{\calF}{{\mathcal F}}

\newcommand{\calK}{{\mathcal K}}

\newcommand{\calM}{{\mathcal M}}

\newcommand{\calS}{{\mathcal S}}

\renewcommand{\emph}[1]{\textit{#1}}
\newcommand{\ogets}[1]{\leftarrow\hspace*{-4pt}#1\:}
\newcommand{\gfloor}[3]{\llfloor#1\rrfloor^{#2}_{#3}}
\newcommand{\gceil}[3]{\llceil#1\rrceil^{#2}_{#3}}

\newenvironment{newmath}{\begin{displaymath}%
\setlength{\abovedisplayskip}{4pt}%
\setlength{\belowdisplayskip}{4pt}%
\setlength{\abovedisplayshortskip}{6pt}%
\setlength{\belowdisplayshortskip}{6pt} }{\end{displaymath}}

\newenvironment{newequation}{\begin{equation}%
\setlength{\abovedisplayskip}{4pt}%
\setlength{\belowdisplayskip}{4pt}%
\setlength{\abovedisplayshortskip}{6pt}%
\setlength{\belowdisplayshortskip}{6pt} }{\end{equation}}

\newcommand{\bnm}{\begin{newmath}}
\newcommand{\enm}{\end{newmath}}
\newcommand{\bne}{\begin{newequation}}
\newcommand{\ene}{\end{newequation}}

\newcommand{\authnote}[2]{\ifnum\authnotes=1\begin{quote}\textbf{#1 says:} #2\end{quote}\fi}
\newcommand{\authfnote}[2]{\ifnum\authnotes=1\footnote{\textbf{#1 says:} #2}\fi}

\renewcommand{\paragraph}[1]{\vspace{.6em}\noindent\textbf{#1}\hspace*{.5em}}

\newcommand{\paragraphb}[1]{\vspace{.2em}\noindent\textbf{#1.}\hspace*{.5em}}

\newcommand{\cmark}{\ding{52}}
\newcommand{\xmark}{\ding{56}}

\newcounter{mynote}[section]
\newcommand{\notecolor}{blue}
\newcommand{\thenote}{\thesection.\arabic{mynote}}
\newcommand{\nsnote}[1]{\ifnum\authnotes=1\refstepcounter{mynote}{\it \textcolor{\notecolor}{(NS~\thenote: { #1})}}\fi}
\newcommand{\tsnote}[1]{\ifnum\authnotes=1\refstepcounter{mynote}{\it \textcolor{cyan}{(TS~\thenote: { #1})}}\fi}

\newcommand{\lbnote}[1]{\ifnum\authnotes=1\refstepcounter{mynote}{\it\textcolor{\notecolor}{(LB~\thenote: { #1})}}\fi}
\newcommand{\vbnote}[1]{\ifnum\authnotes=1\refstepcounter{mynote}{\it\textcolor{purple}{(VB~\thenote: { #1})}}\fi}
\newcommand{\jhnote}[1]{\ifnum\authnotes=1\refstepcounter{mynote}{\it\textcolor{orange}{(JH~\thenote: { #1})}}\fi}
\newcommand{\smnote}[1]{\ifnum\authnotes=1\refstepcounter{mynote}{\it\textcolor{orange}{(SM~\thenote: { #1})}}\fi}
\newcommand{\fixme}[1]{\ifnum\authnotes=1\textbf{\textcolor{red}{[FIXME: #1]}}\fi}

\newcommand{\squishlist}{
 \begin{list}{$\bullet$}
  { \setlength{\itemsep}{0pt}
     \setlength{\parsep}{3pt}
     \setlength{\topsep}{3pt}
     \setlength{\partopsep}{0pt}
     \setlength{\leftmargin}{1.5em}
     \setlength{\labelwidth}{1em}
     \setlength{\labelsep}{0.5em} } }

\newcommand{\squishlisttwo}{
 \begin{list}{$\bullet$}
  { \setlength{\itemsep}{0pt}
    \setlength{\parsep}{0pt}
    \setlength{	opsep}{0pt}
    \setlength{\partopsep}{0pt}
    \setlength{\leftmargin}{2em}
    \setlength{\labelwidth}{1.5em}
    \setlength{\labelsep}{0.5em} } }

\newcommand{\squishend}{\end{list}  }

\makeatletter
\newcommand*\mysize{%
  \@setfontsize\mysize{7.5}{9.0}%
}

\newcommand{\inthesis}[1] {}

\mycomment{

}

\newcommand{\ignore}[1]{\if{0} #1 \fi}
\let\oldnl\nl% Store \nl in \oldnl
\newcommand{\nonl}{\renewcommand{\nl}{\let\nl\oldnl}}% Remove line number for one line
\usepackage{algorithmic}
\usepackage{graphicx}
\usepackage{textcomp}
\usepackage{xcolor}

\DeclareMathAlphabet{\mathcal}{OMS}{cmsy}{m}{n}
\newcommand{\Enc}{\mathsf{Enc}}
\newcommand{\Dec}{\mathsf{Dec}}
\newcommand{\Encode}{\mathsf{Encode}}
\newcommand{\Decode}{\mathsf{Decode}}

\newcommand{\concat}{\,||\,}

\definecolor{orange}{RGB}{255,127,80}
\definecolor{darkgreen}{RGB}{50,127,0}
\definecolor{Blue}{RGB}{0,0,255}

\newcommand\blfootnote[1]{%
  \begingroup
  \renewcommand\thefootnote{}\footnote{#1}%
  \addtocounter{footnote}{-1}%
  \endgroup
}

\AtBeginDocument{%
  \providecommand\BibTeX{{\rm B\kern-.05em{\sc i\kern-.025em b}\kern-.08em
    T\kern-.1667em\lower.7ex\hbox{E}\kern-.125emX}}}

\begin{document}
%-------------------------------------------------------------------------------

\date{}

\title{Leveraging Generative Models for Covert Messaging: \\Challenges and Tradeoffs for ``Dead-Drop'' Deployments}
\subtitle{\huge (Full Version)\footnotemark} %%% 

\author{Luke A. Bauer}
\email{lukedrebauer@ufl.edu}
\affiliation{%
  \institution{University of Florida} 
  \city{Gainesville, FL}
  \country{United States}
}
\orcid{0000-0002-5740-4386}

\author{James K.\ Howes IV}
\email{james.howes@ufl.edu}
\affiliation{%
  \institution{University of Florida} 
  \city{Gainesville, FL}
  \country{United States}
}
\orcid{0009-0007-4476-8125}

\author{Sam A.\ Markelon}
\email{smarkelon@ufl.edu}
\affiliation{%
  \institution{University of Florida} 
  \city{Gainesville, FL}
  \country{United States}
}
\orcid{0009-0004-0968-6165}

\author{Vincent Bindschaedler}
\email{vbindsch@cise.ufl.edu}
\affiliation{%
  \institution{University of Florida}
  \city{Gainesville, FL}
  \country{United States}
}
\orcid{0000-0002-3066-7354}

\author{Thomas Shrimpton}
\email{teshrim@ufl.edu}
\affiliation{%
  \institution{University of Florida}
  \city{Gainesville, FL}
  \country{United States}
}
\orcid{0000-0001-8131-5634}
\begin{abstract}
%-------------------------------------------------------------------------------
State of the art generative models of human-produced content are the focus of many recent papers that explore their use for steganographic communication.  In particular, generative models of natural language text.  Loosely, these works (invertibly) encode message-carrying bits into a sequence of samples from the model, ultimately yielding a plausible natural language covertext.  By focusing on this narrow steganographic piece, prior work has largely ignored the significant algorithmic challenges, and performance-security tradeoffs, that arise when one actually tries to build a messaging pipeline around it.  We make these challenges concrete, by considering the natural application of such a pipeline: namely, "dead-drop" covert messaging over large, public internet platforms (e.g. social media sites). We explicate the challenges and describe approaches to overcome them, surfacing in the process important performance and security tradeoffs that must be carefully tuned. We implement a system around this model-based format-transforming encryption pipeline, and give an empirical analysis of its performance and (heuristic) security.

\end{abstract}

\settopmatter{printacmref=false}
\maketitle

\addtocounter{footnote}{-1}% %% to fix the footnote numbering.

%-------------------------------------------------------------------------------

%-------------------------------------------------------------------------------
\section{Introduction}\blfootnote{*This is the full version of the ACM CODASPY 2024 article with the same title and authors. This version provides additional information about our proposed construction and its empirical performance.}
The state of the art in generative models for natural language text has advanced considerably in past few years, leading to applications across various domains.  
One such application, considered in a number of recent papers, is using machine-learned generative models to create realistic \emph{covertexts}, into which covert message data is embedded (e.g.,\cite{Sallee:2004,Le:2007,Liskiewicz:2011,dai2019towards,Ziegler:2019,Shen:2020,zhang2021provably,kaptchuk2021meteor,ding2023discop}). Informally, the core recipe for such language model steganography schemes is to view the model as a collection of context-dependent distributions over language \emph{tokens} (e.g., individual letters, syllabic blocks, full words, punctuation), and encode covert bits into a sequence of samples from these distributions. To support the realism of the resulting natural-language strings, the sequence of distributions is informed by the history of what has been sampled. Sampling tokens is done in an invertible way, i.e., the covert bits can be recovered from the chosen covertext through an appropriate decoding process.

Producing a covertext in this way, compared to using encryption or traditional steganography that embeds data into an existing cover, ensures that the covertext lacks any obvious distortion, as it appears to be natural text. And although the (cover)text is produced by a language model --- and these models are constantly being improved --- there is mounting evidence that the text they produce is difficult to distinguish from human-written text~\cite{gpt3blogfool,clark2021all}. This allows users to exchange covert messages in plain sight, such as a social media platform, without the adversary being aware any exchange is taking place. This may allow evading even the most oppressive adversaries, for which the mere act of using encryption may be enough to draw suspicion.

Research on such model-based steganography has thus far almost exclusively focused on optimizing the encoding and decoding processes to maximize \emph{capacity and imperceptibility}. Capacity, the average number of covert bits that can be carried in a covertext token, is a core operational metric that has direct bearing upon the lengths of covertexts, the need for fragmentation, etc.  Imperceptibility, the degree to which encoding bits distorts the distribution of samples, has become the de facto ``definition'' of security.  Intuitively, if the distortion is large enough, one might distinguish between covertexts that carry covert bits, and ``covertexts'' that result from sampling with independent random bits.  We argue that this is an extremely narrow way to look at the security of model-based steganography in deployment.

Despite the growing body of papers on model-based steganography \cite{Sallee:2004,Le:2007,Liskiewicz:2011,dai2019towards,Ziegler:2019,Shen:2020,zhang2021provably,kaptchuk2021meteor,ding2023discop}, there has yet to be a serious exploration of applying these schemes to their natural use case: covert messaging in dead-drop deployments. \emph{How should one turn a core of model-based steganography into practically useful covert-messaging channels?}  There has been no real effort to characterize the challenges that arise when one attempts to do this, let alone guidance on addressing these challenges.

Instead, the recent literature seems focused on delivering model-based steganography with incremental improvements in imperceptibility and capacity. 

In this paper, we focus on \emph{model-based covert messaging} via large Internet platforms that serve as dead-drops.  Roughly speaking, we consider a processing flow whereby Alice can (cryptographically) turn a plaintext message into a natural-language covertext, and then post this on a platform (e.g., X, Facebook, Github) for Bob to later retrieve.  
The security intuition for this approach is clear.  There are many millions of posts each day to big social media platforms, coming from nearly as many accounts; meanwhile, large language models can generate posts that look like those created by human platform users. If covert bits are carefully encoded into a model-generated post, finding a covertext becomes a problem of finding a needle in a stack of needles. 

That said, \emph{realizing} this compelling approach to covert messaging is not straightforward. Even putting aside higher-level questions about code distribution, key establishment, and management, etc., and focusing only on end-to-end processing of messages, there are major challenges to address. We briefly highlight some of them.

\paragraphb{Covertext discovery/recovery}
User accounts/rendezvous locations cannot be agreed upon without creating a potential vulnerability that an adversary could exploit. Thus when Alice sends messages to Bob, via one or more covertexts written to the dead-drop platform, Bob faces the non-trivial task of determining \emph{which} of the millions of daily platform messages he should scrape and process. Hence any usable system must support mechanisms for this that are efficient \emph{for Bob}, but not for the adversary.

\paragraphb{Reliability in the face of ambiguity}
Large social media platforms are designed to provide reliable ingest and ``delivery" of posts, but end-to-end reliability means that receiver-side processing should recover plaintexts from covertexts with probability as close to one as possible. However, language model vocabularies are not prefix-free; as a result, the natural-language strings they generate are parsable in myriad ways, preventing reliable decoding. Yet prior work on model-based steganography is silent on methods for reliable recovery. 
Dealing with this parsing ambiguity ---~in a way that is efficient, deployable, and secure~--- may be the most significant challenge for model-based covert messaging.

\paragraphb{Cross-device discrepancies}
It is traditionally assumed that identical computations performed at the endpoints yield identical results. Unfortunately, this is \emph{not} a safe assumption for model-based steganography using large language models across different hardware/software stacks. 
 
Hardware non-determinism and device-specific idiosyncrasies of floating point operations can cause model discrepancies \cite{gundersen2022sources}. Which results in Alice and Bob having different views of the \emph{same} model, preventing accurate bit recovery.

\paragraphb{Platform idiosyncrasies}
Another set of challenges elided in the existing literature involve the specific restrictions that the internet platform enforces upon users' posts. Each platform has its own unique quirks, such as restrictions on the characters that may appear in a post, limits on post length, syntactic embellishments (e.g, hashtags) that support search and categorization of posts, etc. 
Care must be taken to adhere to the restrictions, and to do so in a manner that does not distinguish covertexts from ``normal" post.

\paragraphb{Our contributions}
Any effort to realize the potential of model-based steganography for dead-drop covert messaging will need to address these matters (at least).  Doing so requires decisions ---~largely unacknowledged in prior work~--- about balancing deployment-dependent tradeoffs among efficiency, capacity, and security.
This work aims to surface and explore these challenges and tradeoffs, and to ultimately make these systems practical to deploy.  In particular:
\begin{itemize}[nolistsep,noitemsep,leftmargin=1.25em]
    \item We initiate the study of challenges and tradeoffs %that arise when building 
    of a model-based covert messaging system in a dead-drop deployment scenario.
    \item We propose and evaluate different ways to achieve reliability in message delivery despite parsing ambiguity, cross-device discrepancies, and platform idiosyncrasies.
    \item We propose and evaluate the use of both covert and overt hints to facilitate efficient message discovery and recovery by recipients. 
    \item We describe a security criterion, called plausibility, overlooked by prior work. We also show that systems that fail to ensure plausibility are easily broken by novel trial-decoding attacks.
    \item We discuss and evaluate security beyond imperceptibility (the sole focus of most prior work). 
\end{itemize}

\section{Background and Overview}
\label{sec:background}\label{sec:overview}

\begin{table*}[!t]
    \caption{Summary of security heuristics applied in prior work. \cmark{} indicates the evaluation was conducted whereas \xmark{} indicate that the paper did not mention it or did not evaluate it. }
     \label{tbl:sec}
    
    \renewcommand{\arraystretch}{1.1}
    \centering
    {\small
    \resizebox{.995\linewidth}{!}{
    \vspace{-6pt}
\begin{tabular}{p{1.1in}c c c c c c c c c c c l}\toprule

\multicolumn{1}{l}{} & This Work & \begin{tabular}[c]{@{}c@{}}Ziegler et al. \\  \cite{Ziegler:2019}\end{tabular} & \begin{tabular}[c]{@{}c@{}}Dai and Cai \\ \cite{dai2019towards}\end{tabular} & \begin{tabular}[c]{@{}c@{}}Yu et al. \\ \cite{yu2022mts}\end{tabular} & \begin{tabular}[c]{@{}c@{}}Shen et al.\\ \cite{Shen:2020}\end{tabular} & \begin{tabular}[c]{@{}c@{}}Kaptchuk et al. \\ \cite{kaptchuk2021meteor}\end{tabular} & \begin{tabular}[c]{@{}c@{}}Zhang et al.\\ \cite{zhang2021provably}\end{tabular} & \begin{tabular}[c]{@{}c@{}}Yang et al. \\ \cite{yang2018rnn}\end{tabular} & \begin{tabular}[c]{@{}c@{}}Cao et al. \\ \cite{cao2022generative}\end{tabular} & \begin{tabular}[c]{@{}c@{}}Yang et al. \\ \cite{yang2022semantic}\end{tabular}
&\begin{tabular}[c]{@{}c@{}}de Witt et al. \\ \cite{de2022perfectly}\end{tabular} 
&\begin{tabular}[c]{@{}c@{}}Ding et al. \\ \cite{ding2023discop}\end{tabular}\\  \hline
\parbox{1.1in}{Imperceptibility} & \cmark & \cmark & \cmark & \cmark & \cmark & \cmark & \cmark & \cmark & \cmark & \cmark & \cmark & \cmark\\ 
ML Steganalysis & \cmark & \xmark & \xmark & \xmark & \xmark & \xmark & \cmark & \cmark & \cmark & \cmark & \xmark & \xmark \\ %\hline
Human Evaluation & \xmark & \cmark & \xmark & \xmark & \cmark & \xmark & \xmark & \xmark & \xmark & \xmark & \xmark & \xmark \\ \hline\hline

Decoding Attacks & \cmark & \xmark & \xmark & \xmark & \xmark & \xmark & \xmark & \xmark & \xmark & \xmark & \xmark & \xmark\\ %\hline
Model Evaluation & \cmark & \xmark & \xmark & \xmark & \xmark & \xmark & \xmark & \xmark & \xmark & \xmark & \xmark & \xmark\\ %\hline
\parbox{1.0in}{Identifying Users} & \cmark & \xmark & \xmark & \xmark & \xmark & \xmark & \xmark & \xmark & \xmark & \xmark & \xmark & \xmark\\ %\hline
\parbox{1.0in}{Identifying Msgs} & \cmark & \xmark & \xmark & \xmark & \xmark & \xmark & \xmark & \xmark & \xmark & \xmark & \xmark & \xmark \\ %\hline
\bottomrule
\end{tabular}}}
\end{table*}

\subsection{Model-Based Covert Messaging}\label{sec:overview:messaging}

A model-based steganography scheme consists of a generative model (in our case a language model) and matching encoding and decoding procedures. Roughly speaking, the encoding process takes a bitstring as input and samples from the generative model {\em deterministically} based on the input bits. The samples from the models are then concatenated into a covertext. The decoding process takes this covertext as input and uses the generative model's distribution over samples to infer what bits must have been embedded to produce each sample, thereby reconstituting the bitstring.

\paragraph{Language models and sampling.}
In this paper, we use GPT-2 as language model since it is easily accessible and used almost exclusively in prior works (e.g.,\cite{Ziegler:2019,dai2019towards,yu2022mts,Shen:2020,kaptchuk2021meteor,cao2022generative}). However, our observations and discussion apply equally to any other language model, provided that the model is auto-regressive, meaning that it characterizes a distribution over sequences of tokens according to the chain rule of probability. GPT-2 like other language models uses a finite vocabulary of tokens derived using byte-pair encoding (BPE)~\cite{gage1994new,sennrich2015neural}. This vocabulary is not prefix-free, which as we will describe later results in several system challenges.

Given a context prompt or seed, a language model such as GPT-2 produces a set of \textit{logit} scores over the next possible tokens. Before we can sample from the model, the logits are transformed into a normalized probability distribution using the \textit{softmax} function based on a \textit{temperature} hyperparameter $t>0$. If $z_i$ is the score of the $i^{\rm th}$ token, then the probability of choosing this token is proportional to $\exp(z_i / t)$, meaning that low temperature results in a highly peaked distribution, where the largest logit has nearly all of the probability mass. High temperatures (e.g., $10$) result in nearly uniform distributions over the entire token set. Additionally, some common sampling strategies restrict the set of tokens before transformation. For example, top-$k$ sampling ensures only the~$k$ most likely tokens are included, whereas top-$p$ sampling adds tokens, in decreasing order of probability, to the sampling distribution until a total probability mass of~$p$ has been reached.

\paragraph{Encoding and decoding.}
There are several strategies to embed bits into model samples during the encoding process. Existing schemes used a variety of strategies including Huffman coding~\cite{yang2018rnn,dai2019towards}, bins or grouping~\cite{zhang2021provably, yang2022semantic}, search trees~\cite{yu2022mts,cao2022generative}, and arithmetic encoding~\cite{Ziegler:2019,Shen:2020,kaptchuk2021meteor}. Each technique results in slightly different embedding capacity and some techniques distort the model's natural distribution by sampling only approximately from the token distribution. As a result, a significant focus in prior work is optimizing the encoding process to minimize the distortion. Further, it is argued that higher distortion provides advantages to an adversary and therefore schemes should strive for {\em imperceptibility}, usually measured as the KL-divergence between the effective encoding distribution and the model's natural token distribution.

The decoding process that the receiver must use to recover embedded bits from the chosen covertext token sequence relies on the token distribution probabilities to infer what bits were
embedded in each token choice. However, from the covertext there is no way for the receiver to know for sure what tokens were specifically chosen during encoding, since the model's vocabulary is not prefix-free. Prior works largely ignore this issue and in implementation simply follow the tokenizer's deterministic and greedy token choices, which result in unrecoverable bits for the receiver when different token choices were made during encoding. We discuss later how to overcome this issue. In our case we use arithmetic coding~\cite{Rubin:1979,Howard:1992} as the basis of our encoding and decoding processes.

\paragraph{Cryptographic processing.}
A covert messaging system needs to provide confidentality, authentication, and integrity through cryptographic means. Most prior works do not describe any specific cryptographic process alongside with the encoding and decoding process, although some appear to assume that the bitstream fed to the encoder~\cite{dai2019towards,cao2022generative,yu2022mts,yang2018rnn} is encrypted or uniformly randomly distributed~\cite{Ziegler:2019,Shen:2020,zhang2021provably}. 

In our case, we use a carefully-designed cryptographic {\em record layer} to encapsulate plaintexts. For this we use an Authenticated Encryption with Associated Data (AEAD) construction realized in practice using a block-cipher in CTR-mode and we explain how this construction enables fragmentation of plaintext, which is necessary due to platform restrictions on covertext length.

\paragraph{Record construction.}
We add several features to the record layer that allows us to address challenges later. 
We first prepare the plaintext~$M$ by (1) prepending its byte length as a 1-byte integer, and (2) inserting a distinguished checkpoint byte after each $x$ bytes (in our implementation $x=10$). Both of these features are used to ensure reliable and efficient parsing as well as correct recognition of authenticated messages (as discussed in~\cref{sec:challenges:parsing}). Call this augmented plaintext $M'$.

For the cryptographic operations we derive three keys $K_1,K_2,K_3$ from the shared secret, and require a synchronized message counter ($IV$) as well as a freely chosen \emph{covertext tweak} ($IX$) which allows us to retry covertexts. We encrypt $M'$ using AES in CTR-mode with~$\mathit{IV}$ and $K_3$ to obtain ciphertext~$C$. We also generate a \emph{sentinel value} (SV) by computing $\mathrm{HMAC\mbox{-}SHA512}_{K_1}(``\mathrm{SV}"\concat\mathit{IV}\concat\mathit{IX})$ and truncating the result to a two-byte value~$V$. The sentinel value is our covert hint which we use to facilitate the retrieval of covert messages from the platform for the receiver as described in~\cref{sec:challenges:retrieval}. Finally, we generate the authentication tag $T$ by computing $\mathrm{HMAC\mbox{-}SHA512}_{K_2}(``\textrm{Tag}"\concat V\concat C)$ and truncating the result to five bytes. The \emph{ciphertext record} is the string $V\concat C\concat T$. Again, the choices of two bytes for the SV, one byte for the plaintext length, and five bytes for the tag are to balance security and efficiency, and are adjustable. Note in cases where fragmentation is needed, we include a fragment index at the end of the plaintext chunk instead of the full $T$, as described in~\cref{sec:challenges:platform}.

As mentioned above, we add a message tweak ($\mathit{IX}$) to the construction of the SV. This is a small value, e.g. $0 \leq \mathit{IX} \leq 5$, that allows the user to generate multiple distinct covertexts for the same plaintext. We find this is useful for cases where the encoding method chooses an especially strange covertext, or if the user chooses to simply retry messages instead of preventing errors. Since the space of values is kept small, the receiver can just precompute all expected sentinel values and very quickly screen incoming message by checking for these values.

\paragraph{Rendezvous.}
For the sender and receiver to properly communicate they must pre-share certain necessary information. 
In previous work, this is rarely mentioned, with only a few papers\cite{kaptchuk2021meteor,yu2022mts} stating they exchange model states, so that they are observing identical prediction distributions. 
 We assume our users exchange this along with an overt signal as described in Section \ref{sec:challenges:retrieval}. While this information could be exchanged on a per-message basis, we assume a key table protocol.  
In which, after a one time exchange of a key, and by maintaining a message counter, both the sender and receiver could use a PRNG to select from an identical table of model states. The model state is made up of information such as finetuning, sampling parameters (top-k/top-p, temperature), and an initial seed, anything necessary to synchronize model output. It is important to keep sets of this information because they combine to help the model generate text that looks natural within the context, while also helping the sender and receiver coordinate the exchange location. For instance, such a table would allow the sender and receiver to input a key and message counter, and receive an entry telling them that the current message they are exchanging, would be on the news tag, use a seed of ``In Today’s'' News and be generated with a gpt-2 model finteunted on \#NEWS posts. If they are varied it could also contain sampling parameters such as temperature, and top-k/top-p. 

Here it is worth pointing out Collage~\cite{Burnett:2010}, an end-to-end covert messaging system that utilizes dead-drop platforms that host user-generated content. Collage organizes a covert messaging system into logical layers that resemble the layers in a traditional protocol stack. It also discusses the rendezvous problem (and possible approaches) extensively.

\paragraph{A note about key exchange.}
Any record module implementation is going to require a shared key. To minimize key exchange requirements, in practice it is convenient to derive the record module key from a long-term shared key which is established before communication begins. How this key exchange occurs is not considered here, but we know of at least one approach (\textit{MoneyMorph}~\cite{MinaeiMK20}) which may enable key exchange without requiring a secure out-of-band connection. More generally, steganographic key exchange is a long-standing problem in the field which we do not attempt to solve; the reader may refer to Ker et al.~\cite{Ker:2013} for a more thorough discussion.

\subsection{MBFTE} 
In the rest of the paper, we systematically discuss various challenges that arise when designing a system around a model-based steganographic scheme.  These challenges are not just engineering problems. They require systematic thinking about how to design a system around the steganographic core and navigating the tradeoffs between system performance and security that emerge from various design choices. 

To help us discuss these challenges and evaluate proposed solutions, we implement a potential system  called MBFTE (model-based format-transforming encryption).We stress that MBFTE is not presented as a concrete deployable system nor the only possible solution to covert messaging, but as a tool we use to discuss and analyze performance and security tradeoffs. We show the performance (capacity, encoding/decoding time, etc.) of our scheme in~\cref{sec:performance}.

Here, we provide a technical definition of model-based FTE, the core of our covert messaging system, and then detail a particular construction of an MBFTE scheme. First, let us establish some notation that will be used throughout the remainder of the paper.

\paragraph{Notational preliminaries.} When $X,n$ are integers, we write $\langle X\rangle_n$ to denote the $n$-bit string that encodes $X$. When $a,b,c$ are integers, we write $\gceil{a}{b}{c}$ and $\gfloor{a}{b}{c}$ as shorthand for $(a\cdot2^{c})\:\%\:2^b$ and $\lfloor(a\:\%\:2^b)\cdot2^{-c}\rfloor$, respectively (we call them \emph{bounded bit shifts}). When~$X,Y$ are strings, we write $X \concat Y$ for their concatenation and $|X|, |Y|$ for their lengths. We write $X[i]$ for the $i$th symbol in $X$, $Y[-i]$ for the $i$th-to-last symbol in $Y$, and $X[:i]$ for the string consisting of the first $i$ symbols in $X$.

We use standard pseudocode to describe algorithms, with a few expressive embellishments: When $\star$ is a binary operator, the statement $a\ogets{\star}b$ is equivalent to $a\gets a\star b$. We use the statement $a\ogets{\lambda x.}\Phi$ in a similar manner, evaluating expression $\Phi$ with $a$ in place of each $x$, and assigning the result to $a$. If multiple comma-separated variables appear on the left side of such an assignment, then the operation is applied to each in turn. The expression $\$(n)$ uniformly samples an integer between 0 and $n-1$.

\paragraph{Model-based formats and FTE schemes.}
A \textit{model-based format} is a tuple $\calM=(\Sigma,\calS,\calF,\mathsf{Next},s_0)$ where $\Sigma$ is a set of \textit{tokens}, $\calS$ is a set of \textit{model seeds}, $\calF=\{f_s\}_{s\in\calS}$ is an ensemble of distributions over values in $\Sigma$, $\mathsf{Next}:\calS\times\Sigma\rightarrow\calS$ defines a \textit{seed transition function}, and $s_0\in\calS$ is the initial seed.
A \emph{model-based FTE scheme} (MBFTE) is a pair of algorithms $(\Enc,\Dec)$ with the following specification:
\begin{itemize}[leftmargin=1.2em,noitemsep,nolistsep]
    \item The deterministic encryption algorithm ($\Enc$) takes as inputs a key~$K$, an initial value~$N$, a model-based format $\calM$, and a plaintext string~$M$; it outputs a ciphertext string $X\in\Sigma^*$. We write $X \gets \Enc^{N,\calM}_K(M)$.
    \item The deterministic decryption algorithm ($\Dec$) takes as inputs a key~$K$, an initial value~$N$,  a model-based format~$\calM$, and a ciphertext string~$X$; it outputs a plaintext string~$M$ or the distinguished error symbol $\perp$.  We write $M \gets \Dec^{N,\calM}_K(X)$.
\end{itemize}
\noindent
An MBFTE scheme is \emph{$\delta$-correct} if for all~$N,\calM$, and~$M$ we have $\Pr[\Dec^{N,\calM}_K(\Enc^{N,\calM}_K(M)) = M]\geq\delta$ with probability taken over the choice of $K$; for sufficiently small values of $\delta$ we may simply refer to the scheme as \emph{correct}. These two definitions support formatted encryption schemes for a particular class of generative model; specifically, those which can deterministically produce a family of arbitrarily long distribution sequences from a starting seed.

\paragraph{Sampling as source decoding.}
To transform a plaintext string into a formatted ciphertext, we utilize the same technique as previous work on model-based steganography: encrypt the string, interpret this ciphertext as a source code for the distribution provided by the model, and ``decompress" it using standard source-coding algorithms. This procedure is more nuanced than it may appear because source-coding algorithms are designed in the \textit{forward} direction---given an input string, an optimal encoding for that string will uniquely decode to the same string---but model-based FTE applies these algorithms in the reverse order.

The key challenge is that the ciphertext may not line up with any particular discrete source code implied by the model distribution(s), resulting in ambiguity at the end of the coding process. We pad the ciphertext with extra bits to deal with this.

\paragraph{MBFTE using arithmetic coding.}
The system is parameterized by a symbol length $r$, which denotes the bit length of each symbol (e.g. $r=8$ for byte strings) and coding length $\ell$ which determines the size of the coding range in symbols. The precision of the coding state is therefore fixed at $r\cdot\ell$.

The core of both encoding and decoding is a loop which (1) adjusts the coding range based on the current token, (2) rescales the coding range to shift out determined symbols, and (3) updates the model seed. When rescaling, the range may need to be "inverted" if it straddles a symbol boundary; $w$ keeps a count of symbols that were output in an inverted state, and lines 24--25 adjust these symbols when the inversion is resolved.

During decoding the next token is selected from the model distribution according to the ciphertext bits that are in the coding window (represented by $c$). As symbols are shifted out and appended to $D$, the remaining ciphertext symbols are shifted into the coding window one by one. Once the end of the ciphertext is reached, random padding bits are used to maintain uniform sampling of tokens. The loop terminates once the full length of the ciphertext has been shifted out, at which point $D$ should be equal to $C$, or at least within a small margin of error depending on the value of $w$ and how many padding bits (if any) were shifted out on the final token.

Encoding proceeds in a similar manner, except that the sequence of tokens is provided as input rather than sampled from the model. This parallel operation ensures that, as long as the model $\calM$ is identical, the value of $D$ returned by the encoder will be the same value produced by the decoder. We must also return $w$ because the trailing $w$ symbols will have two alternatives depending on how the inversion could resolve, so both must be considered.

Constructing an MBFTE scheme is now a straightforward composition of the arithmetic coding algorithms with a deterministic AE scheme $\Pi=(\calE,\calD)$ (with keys $K\in\calK$ and IVs $N \in \mathcal{N}$) as follows:
\begin{itemize}[leftmargin=1.2em,noitemsep,nolistsep]
    \item $\Enc^{N,\calM}_K(M)$: Output the value returned by $\Decode(\calE^N_K(M),\calM)$.
    \item $\Dec^{N,\calM}_K(X)$: Compute $C,w\gets\Encode(X,\calM)$. Then,
    \begin{enumerate}[leftmargin=1.2em,noitemsep,nolistsep]
        \item Compute $M\gets\calD^N_K(C)$. If $M\neq\perp$, output $M$ and halt.
        \item If $w>0$ then let $C'\gets C$ and compute $C'[-w]\ogets{+}1$. If $w>1$, invert the bits in the last $w-1$ symbols of $C'$. Compute $M\gets\calD^N_K(C')$. If $M\neq\perp$, output $M$ and halt.
        \item Subtract 1 from $w$, truncate the last symbol from $C$ and return to (1), unless $\ell-1$ symbols have already been truncated, in which case output $\perp$ and halt.
    \end{enumerate}
\end{itemize}
In general, the worst-case number of trial decryptions required\footnote{With suitable values of $r$ and $\ell$ this number can be minimized in practice during honest operation} is $2\ell$. But by ensuring that the message length $|M|$ is encoded into every encrypted plaintext, as we will do, trial decryption can be avoided altogether.

\subsection{Threat Model \& Security}

\paragraphb{Threat model}
The adversary observes posts on the platform, with the goal of identifying those posts that contain covert messages and (or) identifying users of the covert messaging system. We assume the adversary could potentially know every aspect of the system, except the secret key associated with any pair of communicating users. For example, the adversary may have knowledge of the cryptographic scheme used, generative model weights and sampling techniques. 

The adversary may even have access to the system directly, meaning they can act like a user to post messages or attempt to retrieve/decode covert messages from platform posts. 

However, the adversary is not in control of the platform, meaning they cannot prevent posting on the platform or modify existing posts. They also cannot directly see fine-grained actions by users on the platform or directly obtain identifiers (e.g., IP addresses) of those users posting or scraping the platform. Similarly, we assume that banning usage of the platform or taking it down is considered too costly in terms of collateral damage. Finally, the covert messaging system is intended for use on a platform with a significant daily volume of posts (e.g., hundreds of millions per day), so any detection method must scale to be a viable strategy (e.g., human detection would not scale).

Although we do not directly consider it in this paper, it is possible that the platform would cooperate with the adversary. In this case posts could be altered, and the adversary could more closely monitor user activities. Our record layer and sending redundant messages could largely mitigate the effect of the first issue, however the second would require more precaution on the user's side to hide their anomalous actions. We leave this scenario to future work.

Note: our threat model is more comprehensive than that considered by most prior work, which (implicitly) considers an adversary focused on detecting covertext from other text output from the model. Their assumed adversary does not have full knowledge of the system (except the key), access to the system, or the capability to identify covertext by distinguishing it from normal (i.e., human-written) platform posts.

\paragraphb{A note on formal notions} 
While there are security notions for steganography that may apply to our considered  setting~\cite{Hopper:2002,Hopper:2008,Cachin:1998,Zollner:1998,Mittelholzer:1999,Katzenbeisser:2002,Cachin:2004,howes2022security} operationalizing them is challenging. For example, the Hopper et al. notion~\cite{Hopper:2002} requires that the scheme's output be indistinguishable from that of an oracle for the ``true'' channel. The problem is then justifying the choice of the true channel with respect to real-world deployment of the system. One could of course simply assert that the true channel is the underlying language model (essentially defining the problem away as some prior work does) and then prove indistinguishability with respect to that. But this would be vacuous for many deployment scenarios.

\paragraphb{Security considerations in prior works}
Prior work focuses security analysis on imperceptibility. But it is not obvious how imperceptibility relates to security in the deployment scenario we consider --- dead-drop communication over large Internet platforms --- where covert messages are hidden among a sea of ``normal'' platform posts. Identifying posts containing covert message requires distinguishing between the distribution of covertext posts and regular platform posts, which is a measure of {\em naturalness} of the covertext. By contrast, imperceptibility measures only {\em distortion} introduced by embedding covert bits. Another issue with imperceptibility is that measuring it requires using the model's text distribution without embedding covert bits as a reference. But in our considered deployment scenario, hardware and software heterogeneity of devices means there is no single ``reference'' distribution, so imperceptibility measured by Alice may be different than measured by Bob.

\paragraphb{Security beyond imperceptibility}
So if imperceptibility is inadequate to measure security what else should be used? As mentioned this space lacks a proper security definition against which we could evaluate the system as a whole. Nevertheless we can empirically measure detectability of covertexts by evaluating naturalness using a set of machine learning-based distinguishers. We explain how to do this in the paper and evaluate the success rate of various approaches representing different scenarios. It is critical to observe, however, that detectability measured this way reflects the ability of the underlying language model to generate text that matches content on the platform. Newer and more complex language models are more capable than older simpler models, but since almost all prior work uses GPT-2, we also use it. 

Finally, there is another aspect of security that was overlooked in prior work: {\em plausibility}. The model-based covert messaging system must be able to produce the text of every post on the platform. Otherwise an adversary can attempt a ``trial-decoding'' attack on any platform post and if the trial decoding fails conclude that the post is not a covertext. In fact, to ensure plausibility the system must be such that a trial-decoding attack on any post (whether it contains covertext or not) must result in a bitstring indistinguishable from ciphertext bits for any adversary without knowledge of the secret key.  Surprisingly, plausibility is easily violated by certain seemingly unrelated design choices (e.g., to handle or ignore parsing ambiguity) or by using best practices for model sampling methodologies. Numerous prior work's model-based covert messaging schemes violate plausibility and are swiftly broken by one or more variants of the trial-decoding attack. Ironically some of the advice from prior work on how to improve imperceptibility makes schemes more vulnerable to these attacks, not less.

%-------------------------------------------------------------------------------

%-------------------------------------------------------------------------------

\section{Handling Platform Idiosyncracies}\label{sec:challenges:platform}
Each platform has its own unique quirks such as restrictions of tokens it will accept in a post, length limits of posts, and categorization strategy used to organize the posts (e.g., threads, hashtags, etc). Thus any system that uses such platform for covert messaging must contend with these quirks.

\subsection{Length Limits} 
Many platforms have a limit on the length of each post so that only messages of $L$ characters or less are supported. This means that if our input plaintext~$M$ is too long, it needs to be fragmented across multiple covertexts. This is more difficult than it might seem because we do not know \emph{a priori} how to fragment. In particular, we do not know ahead of time how many characters will appear in each model sample, until we actually do the sampling.  Hence, we do not know how many plaintext bits we can encode into model samples before hitting the covertext-length limit. 

For a system construction similar to ours, there are two obvious methods of splitting the message up and sending it across multiple separate platform posts. The first is the simplest, and consists of splitting the plaintext into chunks of a short enough length that one could reasonably assume each chunk will be encoded within the length limit. For instance, MBFTE has a expansion factor of 11.5, with a standard deviation of 2.86 as described in Table \ref{table:1ps}. So for a length limit of 500 bytes (the X character limit), one could assume with significant probability (i.e., 84\%) that a message of 34 bytes will successfully fit on the platform. The downside to this method is that each message chunk would require its own complete record layer, and the sender would need to reject and regenerate covertexts that end up too long. We achieve this by inserting a message tweak in the sentinel value, as described in Section 2.

Our preferred alternative solution is to use our CTR-mode encryption to ``pretend'' as if we \emph{do} know where the fragmentation boundaries are.  We can do this because CTR mode allows us to encrypt the plaintext one bit at a time, in an online fashion.  Thus we can pause encoding of the input when our covertext has neared the covertext-length limit, insert control information (e.g., the fragment index, number of trailing padding bits in the previous fragment) into the plaintext, send the current covertext, and continue on by generating text for the rest of the message. 

Note that when this fragmentation method is employed, one may want to append a sentinel value to each fragment; in particular, when each fragment will result in a distinct platform post. The effective ciphertext record (which is encoded into model samples) would then be $V_1 \concat C_1 \concat V_2 \concat C_2 \concat \cdots \concat V_\ell \concat C_\ell \concat T$, where each $C_i$ is the CTR-mode encryption of a plaintext fragment, each $V_i$ is a sentinel value, and where the tag~$T$ covers everything that precedes it. Including an SV for each fragment does come at a capacity cost that grows linearly in the number of fragments. However it ensures we embed the maximum number of bits in each message, and never need to regenerate covertexts.

We assume a lossless platform however if there were concerns about losing message fragments, giving each fragment its own record layer and sending redundant copies could improve reliability at the cost of efficiency.
Regardless of the chosen method, care should be taken to avoid creating so many fragments that the sender is forced to post a conspicuous number of covertexts to the platform.

\subsection{Token Restrictions}
Some platforms have token restrictions in the sense that specific characters interact with the platform itself (e.g., \# and @ being used to indicate tags and users respectively). A related issue is that some special language model tokens such as ``\textbar endoftext\textbar'' would look out of place as part of a platform post and may act as instant identifiers of covertext. 

If we allow the model to generate such tokens in covertext, the result may be changing the (logical) location of the post or the covertext may be rejected by the platform itself. To account for this, we propose to force the probability of sampling such tokens to 0, effectively preventing them from being produced. Technically, doing so alters the output distribution of the model. However, if the platform itself cannot accept such tokens, or if including them would change how the platform views the post, such tokens would never appear on the platform in the first place. Changing the probabilities for this small set of tokens is trivial and does not measurably alter encode/decode time or capacity. 

Alternatively, the sender could reject and retry any covertexts that generates a forbidden token, however this has the similar effect of altering the base model distribution, and could delay sender-side processing by potentially forcing many retries messages. The amount of retries necessary depends highly on the finetuning and seed being used to generate the covertext, both of which affect the likelihood of a forbidden token being chosen. 

Stepping back, this example highlights the problem with considering security of the steganographic core of the system to the exclusion of the rest. Altering the ``natural'' output distribution of the language model as we advocate here goes against the imperceptibility-maximizing thinking of prior work (see \cref{sec:security:imperceptibility}) --- which proposes increasingly complex embedding schemes to minimize the distance between the language model's natural distribution and that of the produced covertext. But here the distortion introduced actually increases covertness because it removes obvious signs that would otherwise appear on the platform only in covertext posts.
\section{Handling Ambiguous Parsing}\label{sec:challenges:parsing} 
For a model-based covert messaging system to work reliably, the token distributions produced at the sender and receiver at each step of encoding/decoding must match exactly. For language models such as GPT-2, this will be the case provided that the receiver's parsing of a covertext into tokens matches the way the covertext was produced by the sender. Interestingly prior works (e.g., \cite{dai2019towards,Ziegler:2019,Shen:2020}) overlooked this issue, opting the GPT-2's tokenizer as a heuristic for parsing. Although this works in the majority of cases, it fails often enough that receivers cannot retrieve up to 12.5\% of messages intended for them.

\subsection{Nonprefix-free Vocabulary}
The difficulty is that parsing a covertext from the receiver's point of view is inherently {\em ambiguous} because GPT-2's vocabulary (like that of many other language models) is not prefix-free due to the use of byte-pair encoding (BPE)~\cite{gage1994new,sennrich2015neural}. Therefore for some covertexts there are multiple distinct ways to parse it into tokens.

To illustrate this consider the following example. Suppose the first word of the covertext is ``These'', and the tokens `\emph{T}', `\emph{Th}', `\emph{The}', `\emph{These}' all appear in the support of the initial token distribution. From the standpoint of the receiver, the word ``These'' could have been sampled (by the sender) as a single token `\emph{These}' or the sequence of tokens `\emph{The}', `\emph{s}', `\emph{e}', or any other valid parsing of this word into vocabulary tokens. 
Until the receiver commits to a parsing path, decodes the entire covertext, and checks the authentication tag, there is no way to determine \emph{which} of these choices was made by the sender.

Surprisingly none of the prior work on text steganography with GPT-2 mentions this issue. Prior work (e.g., \cite{kaptchuk2021meteor,dai2019towards,Ziegler:2019,Shen:2020}) leans upon the model's native tokenizer ---~a function packaged with GPT-2 which converts natural language into a list of GPT-2 tokens~--- to propose a token path through the covertext. For any given string the tokenizer returns a parsing of it into a sequence of tokens. This method has the advantage of being fast and simple to use since the receiver can run the tokenizer on the covertext and then determine the probability distributions at each step of the proposed path.

However, by generating sets of 1000 covertexts using combinations of parameters described in \ref{sec:performance} and attempting to only decode them using the tokenizer path, we find that for between $3\%$ and $12.5\%$ of covertexts generated with GPT-2 the proposed tokenizer path is not the same as the path chosen by the sender. This rate varies based on the length of the seed, the length of the message, the model fine-tuning, and sampling hyperparameters such as a high top-$k$ or temperature. In deployment, this would be unacceptable as it would mean that up to $12.5\%$ of the messages that Alice sends cannot be recovered by Bob. 

\subsection{Balancing Reliability, Capacity, and Receiver-Side Efficiency}

The common approach is to have the receiver decode each covertext using only the tokenizer path. As previously mentioned, this will fail whenever the sender's sampled tokens do not match the tokenizer path, which means that some valid covertexts cannot be recovered by the receiver. To ensure reliable transmission between Alice and Bob in this case would necessitate a retry mechanism where Alice could regenerate covertexts until one is successfully decodable using the tokenizer path and then send that one to Bob. This would increase the computational burden on the sender, by forcing them to redo the encoding process. For our sample experiments in~\cref{sec:performance}, this would increase encode time by approximately 11.5 seconds per retry. Additionally, it potentially opens up an (admittedly difficult to use) attack vector, because it shifts probability away from some covertexts towards others. For example, if a message is unlikely to be produced through the tokenizer path but nevertheless relatively likely to be produced by the language model, then the adversary can be highly confident that the message is not covertext.

Since this approach is not completely satisfactory in our view, we propose another solution using checkpoint decoding. The solution is motivated by the observation that when the tokenizer path is not the correct one, the actual path is in most cases still relatively similar to the tokenizer path. Concretely, the idea is to use the tokenizer path as the base token path in combination with checkpoints inserted in the ciphertext to verify that the current token path is likely the correct one. When the current path is not correct the receiver process backtracks to the most likely alternative and attempts that path.

The sender inserts single zero-byte checkpoints, into the plaintext record every $x$ bytes, where $x$ is a system parameter that tunes the tradeoff between decoding effort and capacity. Every time a prediction is requested from the model, there will be a token predicted by the tokenizer path, and multiple alternative tokens that could possibly match the covertext chunk being currently parsed. For each of these alternative tokens we record the token, the location in the text, the model context, the current number of backtracks, and the probability compared against the tokenizer predicted token. This record is added to a stack sorted by token probability when compared against the tokenizer predicted token. For the first pass through, we always choose the token selected by the tokenizer. Then the chosen token is added to the list, and we decode as much of the ciphertext as we can using the current token path. 

Once we have enough of the ciphertext to check either the sentinel value, one of the checkpoints, or the final MAC tag, we do so. If the value is as expected, we lock in the current path, reduce likelihood of alternative token choices currently on the stack, and continue on the current tokenizer path.  

If the checkpoint value does \emph{not} match what is expected, we pop the most likely alternative token off the stack and backtrack to it. We set the current location and model context to the values associated with that token. To determine what path to take going forward, we run the tokenizer over the remaining covertext that has not been processed in the current path, and set that as new tokenizer path. We repeat this process, until we pass all the checkpoints and the final MAC tag. 

The downside of this approach is that the single byte checkpoints inserted every $x$ bytes reduce embedding capacity. The advantage is that this decoding strategy is more efficient in practice than alternatives. This is because with overwhelming probability the correct token sequence selected by the sender either matches the tokenizer path or has few differences with it, meaning decoding time is only increased by a small amount compared to the tokenizer decoding described above. Further, because this strategy allows for backtracking until the correct path is identified (no matter how far from the tokenizer path it is), Bob always recovers the message correctly (unlike techniques used in prior work).

\section{Identifying and Retrieving Messages}\label{sec:challenges:retrieval} 

Users cannot simply exchange usernames, since the adversary may intercept it, so the receiver needs a way to identify which posts on the platform contain covert messages intended for them, and do so in a way that the adversary cannot take advantage of. To aid Bob in receiving the intended message, we propose adding covert signals and overt signals.

\subsection{Overt Signaling} 

Overt signals are publicly visible, and do not depend on the cryptographic operations that created the covertext. When possible, we should choose overt signals that can be filtered via the platform API (i.e., without needing to be scraped) such as hashtags on X or Mastodon --- this greatly reduces the number of messages that need to have their covert signals checked. In our implementation, overt signals are hashtags added to the end of the covertext.

Good overt signals must strike a balance between efficiency, capacity, and undetectability, as they determine the amount of content that the receiver must sift through, and may also reveal the presence of covert messages to an adversary. A tag with too many posts makes it difficult for the receiver to find, but too few posts makes identifying covert messages trivial for an adversary with knowledge of the tag. The average post length also dictates how long covertexts can be without becoming trivial to detect.

\begin{figure}[!t]
    \vspace{-6pt}
     \centering
     \includegraphics[width=.925\linewidth]{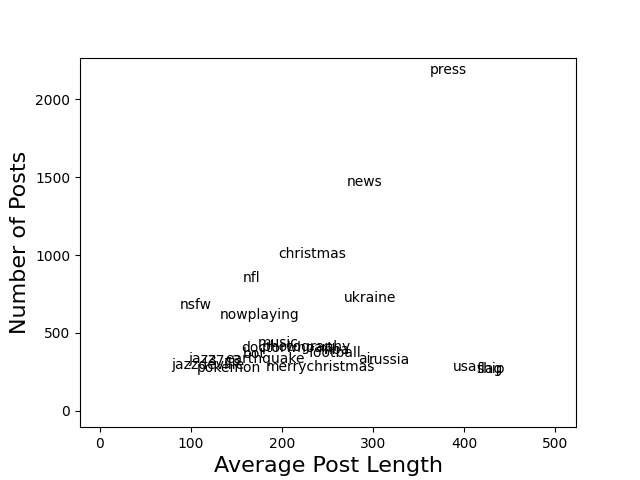}
     \vspace{-4pt}
     \caption{Out of 100k posts scraped from Mastodon.social on December 26th, 2023, we show the 25 most used tags. For each tag, we show average post length against number of posts containing it.}
     \label{fig:toots}
     \vspace{-6pt}
 \end{figure}

In Figure~\ref{fig:toots} we show popular tags and their average post lengths. We scrape 100k posts from Mastodon.social, and compare the average post length and the number of posts for each tag. Each tag offers tradeoffs in detectability and capacity. The ``press'' tag has a relatively high average post length and many posts to hide among, however the high traffic may make retrieval difficult. A tag such as ``immersive\_vr\_experience'', not shown since it is rarely used, averages posts of 490 chars, but only contains 2 posts, making detection trivial. A tag such as 'usa', has about 300 posts averaging 400 characters, and provides security as well as capacity.  

\subsection{Covert Signaling}\label{sec:sentinel} 
Even with well-chosen overt signals, on a large platform, there may be hundreds or thousands of posts to check. To avoid having to fully process each candidate post to see if it indeed contains a message intended for Bob, we add covert signals to each message.

Covert signals are not publicly visible and are part of the cryptographic operations that form the bitstream. Our covert signal is a sentinel value (SV) added to the front of the ciphertext record during its construction. This value is a 2-byte HMAC of the message counter and a message tweak made using the shared keys. The addition of the SV means that during decoding we only need to parse enough tokens to unambiguously determine the first two bytes rather than the entire (potential) covertext. We can then recompute the expected SV and if it matches the recovered two bytes, then the receiver knows that it is likely processing an MBFTE message.

To evaluate the benefit of this approach, we experimentally measure the time for the receiver to decode and check the sentinel values of a collection of 100 messages (99 real Mastodon posts, 1 MBFTE message) when the collection is in random order. Over 50 trials, this process took $57.9$ seconds ($\pm 3.27$ seconds). This result shows that the receiver can periodically poll the platform in order to retrieve new MBFTE messages. Given that Mastodon, Twitter, and Reddit provide functionality for chronologically viewing posts, as well as selecting subsets through hashtags or subreddits, it is not difficult to check all messages in a given time period and semantic signal. Without using covert signaling (i.e. attempting to fully decode every message) the same process takes at least $19.3$ minutes, complicated massively by the fact that checkpointing would fail on all non-MBFTE messages. If the recipient does not know if a message is intended for them, reliable decoding fails since the receiver cannot tell if they are on the wrong path, or the wrong message.

\section{Handling Cross-Device Discrepancies}\label{sec:challenges:cross}
A well-known challenge in machine learning reproducibility is that hardware non-determinism and idiosyncrasies of floating points handling across different devices and architectures
can result in different outcomes of the same computation even when starting from an identical model and identical inputs~\cite{gundersen2022sources}.

In the case of model-based covert messaging this problem manifests itself by the sender and receiver seeing different token distributions from the language model. Prior works implicitly assumed that both sender and receiver would have the exact same view of model outputs, but differences in hardware and software stacks renders this assumption incorrect. Any difference in observed token probability prevents decoding of the message, and thus this issue prevented communication for $100\%$ of messages exchanged between different architectures. 

\subsection{Increasing Floating-Point Precision}

We found that moving from full-precision floating points (32 bits) to 64-bit floating points can alleviate the problem and reduce differences between sender and receiver views to the point that reliable decoding is achieved.\footnote{For example, with PyTorch this can be easily achieved by adding the code line \texttt{torch.set\_default\_dtype(torch.float64)} before loading the model.}

An obvious drawback of this approach is that it substantially increases model inference time and memory consumption (by roughly 20\% and 50\% respectively, in our experiments) and this increase affects both the sender and the receiver. The inference time is likely acceptable, but the increase in memory usage, is a significant hindrance, especially for older mobile devices that may not have sufficient memory to even load the model making the system unusable. It is also worth noting that while we have not observed any failures with 64-bit floating points, we cannot conclusively determine that such failures will not occur for device combinations not tested.

\subsection{A More Principled Solution}
We propose a different solution that leans into the possibility that the sender and the receiver views of model token distribution may be distinct. Two types of errors will prevent the exchange of messages between the sender and receiver. The first type occurs if there is a sufficiently large difference in the observed probability of Alice and Bob for a given token. Recall that the encoding process chooses a token based on the bits it needs to embed according to the sender's token distribution. If the receiver sees different probabilities for the chosen token, it will decode different bits than were encoded. To avoid this, the predicted probability of tokens with embedded bits must be identical. This problem can be solved by rounding token probabilities until they are identical across devices.

The second type occurs when the differences in probabilities between devices is larger than the probability of a token. This causes them to swap positions within the (sorted) distribution, and can cause discrepancies regardless of rounding. To maintain reliability, our idea is to have the sender and receiver agree on a common subset of likely tokens (and their probabilities) that captures most of the probability distribution at each step. We assume a maximum distortion bound $\delta > 0$ on the maximum difference between token probabilities in the sender and receiver views. Subsequent tokens with a likelihood less than $\delta$ are unreliable. With this the sender and the receiver determine the common ``trusted'' subset independently by looking at their own view of the token distribution. If the encoder chooses a token within the trusted subset, then the process continues as usual and bits are embedded. Otherwise, a token is chosen {\em randomly} within the untrusted subset according to the sender's token distribution. In this case the chosen token is still added to the covertext, however no bits are embedded. The receiver can then independently check to see if the observed token was part of the trusted subset, if not it is skipped (since it does not contain any covert bits) and the receiver continues trying to retrieve bits from the next token in the covertext. 

Unfortunately, not embedding bits under certain conditions presents a new problem. Arithmetic encoding determines where in the distribution to select a token from based upon the bits currently being embedded. If those require the encoder to pull from the ``untrusted'' tail of the distribution, there is a possibility that the encoder will get stuck in a loop of repeatedly selecting tokens from the tail, without ever embedding any bits. 

It's worth noting here that a similar phenomenon was observed in other arithmetic embedding schemes from prior work such as Meteor, and de Witt et al.~\cite{de2022perfectly}. They do not use our trusted filter method, but they encounter sequences of high probability tokens that are repeatedly chosen but do not allow for the embedding of bits, leading to a loop with the covertext growing indefinitely without making progress. 

We propose a simple solution to both problems: flip the ordering of cumulative distribution (from ascending to descending and vice versa) at each encoding step, so that each step uses a different ordering than the next. Doing this does not alter the token distribution in any way, since each token takes the same proportion of the distribution as before and arithmetic coding does not care about token order. But this does cause the same embedding bits to select from different parts of the distribution at each step. Since the flipping of the ordering of the distribution is deterministic it is trivial to replicate it on the receiver side.

This process does not influence imperceptibility (KL-divergence). The downside to this solution is that the covertext expansion does increase approximately by a factor of two (depending highly on how much of the distribution is vulnerable to location swaps.)

\section{Performance}\label{sec:performance}

What are the relevant performance metrics for a model-based covert messaging system? 
Two straightforward aspects of system performance are capacity and processing times. 
All else being equal it is of course desirable to maximize capacity. However, capacity is normally constrained by the naturalness (e.g., lack of randomness) of the underlying language model. Further, sampling parameters and encoding strategy also affect capacity. We define sender side processing time as the time to encrypt and encode a plaintext into a covertext and receiver side processing time as the time to decode and decrypt the covertext back to the plaintext. These processing times depend on the plaintext length and bits/token, but also on the design choices made to handle challenges such as handling parsing ambiguity (\cref{sec:challenges:parsing}). This is why two covert messaging systems that rely on the same underlying model may have drastically different processing times. 
\begin{table}[!t]
\centering
    \caption{Performance statistics over 1000 trials. All times reported are measured in seconds.} 
    
    { \small
    \begin{tabular}{|l|r|}
    \cline{2-2}
    \multicolumn{1}{c|}{\ }                          & Mean ($\pm$ Std)   \\ \hline \hline
    
    Capacity per Token      & 3.34  ($\pm$ 0.53) \\ \hline
    Encoded Expansion     & 11.5  ($\pm$ 2.86) \\ \hline
    Plaintext Bits per Covertext Bits    & 0.065 ($\pm$ 0.01)   \\ \hline
    Mean Sender-side Time         & 10.1  ($\pm$ 3.62) \\ \hline
    
     Sentinel Value Check Time         & 0.795 ($\pm$ 0.09)  \\ \hline
     Mean Receiver-side Time                  & 11.58 ($\pm$ 8.05)  \\ \hline

    Receiver-side Failure Rate                        & 0.00   \\ \hline
    \end{tabular}
    } %%
\label{table:1ps} 
\end{table}

\subsection{Measuring Capacity}

We measure bits/token as well as overall encoded expansion for MBFTE, to provide accurate evaluations of how our record construction and checkpoint additions do decrease overall embedding efficiency. For both capacity and processing times measurements we performed a single experiment

We use a machine with an Intel Core i7-6700 CPU with 8 GB of RAM. This machine has no dedicated GPU. While our system would run on less powerful machines, we posit that this machine is representative of the average hypothetical desktop user. 

The seed for all performance experiments is the line from Isaac Asimov: ``\textit{Or maybe it could be put more simply like this: How can the net amount of entropy of the universe be massively decreased?}'', the temperature is 0.8, and 
no top-$k$ or top-$p$, aka the full distribution.  
We measure the capacity per token as the number of encoded bits (plaintext bits plus record layer bits) divided by the number of generated tokens and the encoded expansion is covertext bits divided by encoded bits. Unless otherwise specified, we use the 124M GPT-2 small model with no fine-tuning. Note, that fine-tuning does affect capacity per token since it narrows the model distribution to better fit the data it was trained on. To better show trade-offs, and since the exact influence is highly dependent on the devices in question, we provide the time and capacity measurements for a system without the more complex cross-platform optimization.

\subsection{Parameter Tuning \& Tradeoffs}
We optimized MBFTE's parameters for covertness and reliability. For example we use a temperature of $0.8$ since this is the least detectable temperature we examined. At this temperature we achieve approximately 3.34 bits/token. If we were to increase temperature to $3$, we average $15.07$ bits/token, but the distribution over tokens becomes close to uniform which results in outputting gibberish.

Capacity also depends on the model finetuning used. In experiments we observed that finetuned model typically yields a next token distribution more concentrated on a few tokens than the base model, which makes sense since it is able to generate more specific text. Concretely, our model finetuned on Mastodon \#news posts averages a 1.88 bits/token capacity, but is significantly less detectable when compared against other \#news posts. 

\subsection{Processing and Platform Times}\label{sec:performance:processing}
Processing time is most directly related to the number of token predictions made by the generative model. The shorter the covertext, the less processing time needed. For this reason most prior work do not report processing times, with only two reporting the time it takes their model to generate 50 tokens/words~\cite{cao2022generative,yang2018rnn}. Kaptchuk et al.~\cite{kaptchuk2021meteor} do perform a more in-depth examination of processing times for different hardware and embedding/compression methods. Although hyper-parameters, hardware, and plaintext length all differ between our setups, their base CPU setup reports approximately 80 seconds to encode and decode a 160 byte message. Their compressed method achieves significantly better times, at approximately 40 seconds for an 160 byte message, although the improvement seems to largely comes from generating a much shorter covertext. 

As can be seen in Table \ref{table:1ps}, for a 40 byte plaintext, MBFTE takes approximately 11.5 seconds to encode the plaintext into a covertext and 11.58 seconds to decode the plaintext from the covertext. Note that out of the 1000 messages, approximately $3.4\%$ of them did not follow the tokenizer path, meaning we used checkpoints and backtracking to successfully decode them.

\paragraph{Platform times.}
Sender and receiver processing times only measure encode and decode speed. So to get a more holistic picture of the system's performance we should also consider platform times --- time taken to upload a covertext to the platform or time taken to retrieve covertexts from the platform. As we previously mentioned, none of the existing work considers platform times. We used the Python wrapper for the Mastodon API~\cite{halcy} to post and scrape messages and found that the time to post and scrape is essentially just that of the network round-trip time\footnote{It should be noted that the Mastodon API rate limits by IP address to 300 requests per 5 minutes}. This means that performance is bottlenecked by local operations (encoding/decoding). However, we found a much larger issue is the receiver identifying which message is intended for them. It is for this reason we introduced 
 both covert and overt hints as described in \cref{sec:challenges:retrieval}.

\section{Security Heuristics in Prior Work}
In this section, we discuss heuristics used in prior work. In later sections, we broaden the scope of security heuristics considered. \cref{tbl:sec} provides a summary.

\subsection{Imperceptibility} % : 
\label{sec:security:imperceptibility}
The most common heuristic used by prior work to evaluate security is {\em imperceptibility}, which is often measured as the KL-divergence (KLD) between the output distribution of the system (i.e., that of its produced covertext) and the output distribution of the underlying language model (GPT-2)~\cite{Ziegler:2019,dai2019towards,Shen:2020}. Other prior works also, or alternatively, use perplexity to describe imperceptibility.

Measuring imperceptibility enables straightforward comparisons between various schemes. Since distortion introduced in sampling from the underlying language model may provide an advantage to an adversary, lower KLD values indicate higher imperceptibility and presumably better security. As a result, several papers modify sampling to reduce the KLD compared to the prior art. For example: Dai and Cai~\cite{dai2019towards} only embed bits when the KLD for that step is sufficiently small. Shen et al.~\cite{Shen:2020} dynamically adjusts the distribution size to minimize the KLD. Other sampling strategies, such as arithmetic encoding~\cite{Ziegler:2019}, grouping~\cite{zhang2021provably,yang2022semantic}, and Meteor~\cite{kaptchuk2021meteor}, do not take KLD directly into account during the sampling process, but are still chosen to minimize their effect on the underlying model distribution and thus reduce KLD. \cref{tbl:kld} shows KLD measurements for prior work and our system.

\begin{table}[th!]
    \caption{KLD values presented in this work and prior works. 
    }
     \label{tbl:kld}

    \centering
    {\scriptsize
    \begin{tabular}{c c c c c c c c}
    \toprule
    This Work & \cite{Ziegler:2019} & \cite{dai2019towards} & \cite{Shen:2020} & \cite{kaptchuk2021meteor} & \cite{zhang2021provably} & 
 \cite{de2022perfectly}&
 \cite{ding2023discop}
 \\ \hline
0.061 & 0.0013 & 0.12 & 0.1 & 0.045 & 0.027 & $10^{-16}$ & 0  \\ \bottomrule
\end{tabular}
    } 
    
\end{table}

Measuring security through imperceptibility (implicitly) assumes that the adversary is seeking to distinguish GPT-2's natural output from the system's covertext. But, in many natural deployment scenarios such as the one we consider (i.e., covert communication over large and public Internet platforms) the covertext is observed by the adversary alongside normal messages. In such cases, an adversary that seeks to identify covert communication is instead attempting to distinguish between covertext and normal messages --- not between covertext and sampled GPT-2 text.

\subsection{ML-based Evaluation}\label{sec:security:prior:ml}

Machine learning distinguishers can also be used as a heuristic for evaluating security. 
In this case the adversary is modeled as a binary classifier tasked to distinguish between covertext and normal platform messages. The classification decisions are made based on a threshold, which can be varied based on adversarial objectives between the two types of errors that the adversary can make: false alarms (i.e., false positives) and missed detections (i.e., false negatives).

A small subset of prior works follows this approach to evaluate security. Of these works, only Cao et al.~\cite{cao2022generative} use GPT-2. They use a text steganalysis RNN~\cite{yang2019ts} to distinguish between covertext messages and text from movie and tweet databases. Their detection model achieve a maximum accuracy of about $0.74$. They find that the embedding rate has a strong effect on how detectable their output is. Other works that use machine learning distinguishers \cite{yang2018rnn,yang2019ts,zhang2021provably} use a variety of steganalysis models~\cite{yang2020ts,yang2019fast,peng2021real,meng2009linguistic,chen2008linguistic}, but also use different models to generate text, so their results are difficult to compare.

Aside from the inevitable limitation that measuring security with machine learning only provides a lower bound on (in)security --- because better classifiers may be developed in the future --- existing works are limited in the factors they consider. Of all the works that address ML-based adversaries, the only variable that is examined past the base steganographic model is embedding rate. They do not consider different adversary models, hyper-parameters, or other factors that impact detectability such as fine-tuning. Note that there exist machine learning classifiers specifically designed to detect output from a specific language model (e.g., RoBERTA~\cite{openai-roberta} for GPT-2) but prior work does not use them.

\section{Covertext Plausibility and Decoding Attacks}
A model-based covert messaging system must be able to \emph{plausibly} produce the text of every post on the platform. Otherwise, an adversary can filter out any posts that cannot be produced as not covertexts, potentially making covertext detection trivial.
To demonstrate the problem, we describe two attacks that exploit implausibility with respect to the model and its use; recall that adversaries are assumed to know everything except secret keys. We also show how a covert messaging system can avoid this vulnerability. 

\subsection{Decoding Attacks}\label{sec:security:decoding}
Consider an adversary that attempts to detect covert messages by performing a ``decoding attack'' whereby it attempts to decode suspect platform posts using the receiver-side process. Specifically, an adversary acting as a user of the system, with full knowledge of the model configuration (i.e., model, seed, sampling hyperparameters, etc.) --- just not the secret key --- can scrape posts from the platform and attempt decoding. 

Note that for any post containing covertext, this adversary cannot successfully decrypt the ciphertext to recover the plaintext without the secret key. However, if the decoding behavior is different on regular platform posts than on posts containing covertext, then the adversary can easily differentiate between the two. An example where this may occur is for sampling schemes that restrict the set of possible next tokens such as top-$k$ and top-$p$, or restricted tree based sampling methods~\cite{yu2022mts,cao2022generative}. For every covertext at nearly every $k$ or $p$ value, the model can parse a potential path through the covertext, and therefore decode it. This is not the case for many or most real platform posts because some of their tokens are not within the top-$k$ and top-$p$ so that the decoding process must abort before recovering a complete ``ciphertext'' string.  

Unfortunately, some prior work often uses (relatively) small values of $p$ and $k$ (e.g., $k<1000$ most of the time)~\cite{Ziegler:2019,Shen:2020}, making these schemes particularly vulnerable. When evaluating GPT-2 de Witt et al~\cite{de2022perfectly} performs some experiments with top-k=40, and some with top-p=90\%. Ding et al. ~\cite{ding2023discop}, uses top-p $\leq$ 100\% for many experiments although they do also show results for using the full distribution.  Shen et al.~\cite{Shen:2020} uses a variant of top-$k$ sampling with a dynamically adjusted $k$ to lower KL-divergence.  
Other papers that do not directly rely on top-$k$ or $p$ sampling may still restrict their next token distribution based on other factors, such as Yu et al.~\cite{yu2022mts} who restrict each level of their search tree to approximately $2^6$ possible values, essentially resulting in a top-$k$ of 64. 

\begin{table}[!htb]
  
    \caption{Percentage of decodable real posts on Mastodon \#News for varying sampling strategies and parameters.}
    
    \label{tbl:samplingdec}
    \centering
    \begin{minipage}{.5\columnwidth}
        \centering
        {\small
        \begin{tabular}{|l|l|}
        \hline
        Top-$k$ & \begin{tabular}[c]{@{}l@{}}Real Posts \\Decodable\end{tabular} \\ \hline
        100 & 0\% \\ \hline
        1000 & 70.3\% \\ \hline
        2000 & 95.3\% \\ \hline
        3000 & 100.0\% \\ \hline
        \end{tabular} } %
        %\caption{Top-$k$ Sampling.}
        \label{tbl:topkdec}
    \end{minipage}%
    \begin{minipage}{0.5\columnwidth}
        \centering
        {\small
        \begin{tabular}{|l|l|}
        \hline
        Top-$p$ & \begin{tabular}[c]{@{}l@{}}Real Posts \\Decodable\end{tabular} \\ \hline
        0.9 & 0\% \\ \hline
        0.95 & 9.4\% \\ \hline
        0.99 & 81.3\% \\ \hline
        0.999 & 100.0\% \\ \hline
        \end{tabular} }%
        
        \label{tbl:toppdec}
    \end{minipage}
    
\end{table}
To understand the relationship between the decoding attack success rate and sampling strategies we perform the following experiment. We scraped 1000 posts from Mastodon News and recorded the proportion of them that could potentially be generated by GPT-2 using the base 124M model for different sampling hyperparameters. The results are shown in~\cref{tbl:samplingdec}. The decoding attack has a significant advantage when using small values of hyperparameter $k$ or $p$ (sometimes to the point of a complete break).

However, even when using large values of $k$ (or $p$) such as setting $k=3000$, the adversary still has an advantage. We found that only 8.8\% of real posts could be fully traversed by the tokenizer using tokens within this top-k, whereas every single steganographic message with this $k$ could be fully traversed. This means that an adversary could attempt to decode potential covertexts using the tokenizer path, and deduce that any post that is {\em not} compatible with this path is {\em not} a covertext.

Our proposed system avoids this vulnerability completely by always using the entire token distribution when sampling. Tokens forbidden by the platform, as described in Section \ref{sec:challenges:platform}, can be excluded without creating a vulnerability, since other platform messages cannot contain them either. Stepping back, this vulnerability highlights the fallacy of focusing the security analysis on the output distribution of covertext versus the natural distribution of the model, instead of also accounting for system-level design choices (e.g., how to handle parsing ambiguities) which have profound security implications.

\subsection{Distribution of Recovered Bits}\label{app:recovered-bits-distribution}
If a normal platform post can be trial-decoded successfully, the adversary will recover some bits after decoding it. Consequently, in order to completely thwart decoding attacks, the recovered bits from any trial decoding of normal platform posts have to be indistinguishable from those recovered from covertexts produced by the system. Therefore when designing the system, we need to ensure that the distribution of ciphertext bits produced during encoding matches that of the distribution of bits recovered from trial decoding normal platform posts. 

Prior work seems to largely assume the bits being encoded are the actual message bits encrypted into a uniform distribution \cite{Ziegler:2019,dai2019towards,Shen:2020}, although they refrain from mentioning any specific cryptographic method. The main exception being Kaptchuk et al.~\cite{kaptchuk2021meteor}. 
To see if this threat can be avoided with proper cryptography we performed this attack against our cryptographic layer (\cref{sec:overview:messaging}). Our cryptographic record creation produces a ciphertext bitstream that is uniform. 

We empirically evaluate the distinguishability of bits recovered from normal platform posts versus covertexts for our system using (1) entropy tests and (2) statistical tests of randomness. The idea is that since for MBFTE our ciphertext bits are the output of an AEAD scheme they cannot be distinguished from uniformly random bits. Specifically, we constructed 1000 byte segments by concatenating bitstreams of recovered bits from the decoding of multiple messages for both covertext and real platform messages (scraped from Mastodon News). We then calculated the byte-level entropy of these segments and normalized it. We then repeated this experiment $100$ times to get an average and standard deviation. We measured an average entropy of $0.967$ ($\pm 0.0034$) for real platform posts and $0.971$ ($\pm 0.0026$) for covertexts. Calculating entropy over segments of 1000 bytes provides a reasonably accurate estimate, but it is unrealistic for an adversary to obtain segments of 1000 bytes given the capacity and expansion factor. So we also repeated the experiment using segments of 100 bytes (which better reflects the amount of information an adversary would obtain from a single message). In that case, we observed an average entropy of $0.779$ and $0.784$ for real platform posts and MBFTE covertexts, respectively. We conclude that an adversary is unable to distinguish real messages from MBFTE messages using entropy measurements.

\paragraphb{Plausibility According to Statistical Tests}
\label{sec:plausibility:entropy}
Instead of entropy tests, an adversary may attempt to distinguish recovered bit strings using other statistical tests. To evaluate this, we use the NIST statistical test suite for evaluating pseudorandomness~\cite{smid2010statistical}. This test suite has been used for the design and cryptanalysis of pseudorandom number generators for cryptographic applications. We used the public implementation of the test suite\footnote{https://github.com/terrillmoore/NIST-Statistical-Test-Suite} and ran all the tests on 10 segments of 10,000 bits from each distribution. Bitstreams for both real posts and covertexts passed all the tests, except for the serial test where platform messages only passed 6/10 of the serial tests while MBFTE messages passed 8/10 of the serial tests. We believe this may be due to the way the 10,000 bits segments are constructed (which necessarily come from multiple messages). To further evaluate the scenario where an adversary recovers bits from a single message, we repeated the experiment using segments of 400 bits. In that case, some tests (including serial tests) could not be performed. But both distributions passed all remaining tests.

\section{Security Considerations in Deployment}
In this section, we discuss security considerations beyond those considered in the prior work and plausiblity. These include naturalness of the language model text on the platform, as well as detection of individual messages and covert-messaging users.
 \begin{figure}[!t]
    \vspace{-6pt}
     \centering
     \includegraphics[width=.925\linewidth]{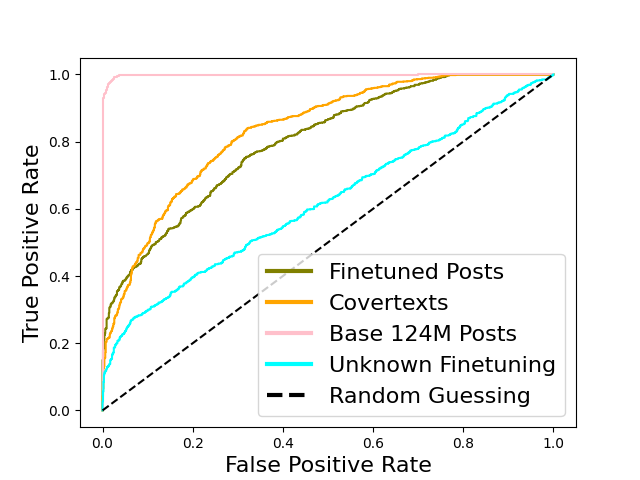}
     \vspace{-4pt}
     \caption{ROC curves for base 124M GPT-2 generated posts, finetuned GPT-2 generated posts, MBFTE covertexts generated with finetuned GPT-2, and finetuned posts where the adversary is unaware of the finetuning being used.  }
     \label{fig:rocs}
     \vspace{-6pt}
 \end{figure}

\subsection{Naturalness of GPT-2 Text}\label{sec:security:ml}\label{sec:security:naturalness}
How natural is GPT-2's output? Can it be easily distinguished from regular platform posts? In this section, we empirically investigate these questions by using an ML-based adversary to distinguish between GPT's text output and platform (Mastodon) messages. These questions are highly relevant to detectability of covertext for model-based covert messaging systems. However, note that what we are evaluating here is not the detectability of covertext but the naturalness of the language model's output on the platform. This is an important distinction because regardless of the imperceptibility of the steganographic technique --- most methods including ours have low KL-divergence with the model's natural output --- if the model's output does not fit the platform, it can readily be detected by an adversary.

We frame the problem as a binary classification task. Given sample text, the task is to predict whether it is a regular platform post or the output of GPT-2. We use a Receiver Operating Characteristics (ROC) curve, which plots true positive rate (TPR) versus false positive rate (FPR) and we calculate the Area Under Curve (AUC).

We use OpenAI's RoBERTa model, and generate 3 sets of $11,000$ GPT-2 messages using seeds gathered from Mastodon \#news, a temperature of $0.8$, and top-$k$ sampling with $k=50000$, which is equivalent to the full distribution. The first set is generated using the base 124M GPT-2 model. The second set is generated using a 124M GPT-2 model that has been finetuned on posts scraped from Mastodon \#news. The final set uses the finetuned model and embeds messages into the generated posts using MBFTE. 

For each set of generated posts, we train a RoBERTa model to differentiate between 10,000 samples of the generated set and 10,000 of scraped real \#news posts. 
We train the models using the default parameters values, which include a $2e^{-5}$ learning rate and a batch size of $24$ and stop when the training accuracy stops increasing (after about 15-20 epochs). Additionally, we evaluate the scenario of the adversary not being aware of the finetuning, and thus attempt to detect finetuned covertexts using a RoBERTa model trained on base 124M generated posts.

From the result in~\cref{fig:rocs} the base 124M GPT-2 output is highly detectable (AUC=$0.998$) due to the fact that news posts have a very specific format and phrasing that base GPT-2 does not imitate well. With finetuned models, we get significantly better output with for finetuned posts without embedded messages (AUC=$0.801$) and with embedded messages (AUC=$0.8332$). Finally, when the adversary does not know the finetuning, they achieve an AUC of $0.618$. 

We can also compare adversaries by fixing the false positive rate (FPR) and considering the true positive rate (TPR). We find that the TPRs for FPR = $0.1\%$ ---~the lowest rate that we observed and can accurately estimate~--- are $90.07\%$ for 124M posts, $15.2\%$ for Finetuned posts, $2.7\%$ for covertexts, and $1.7\%$ when the finetuning is unknown. Low TPR at such low FPR indicate the difficulty in accurately detecting finetuned GPT-2 output and covertext without significant false positives. We expect larger and more advanced language models that exhibit greater naturalness to be even less detectable in the low false positive regime.

\subsection{Detecting Individual Messages} 
An important observation is that detecting covertext posts on the platform is {\em not} the same as classifying a post as normal versus covertext. This is because once the system is deployed, we need to consider the base rate of covertext posts on the platform. The base rate is dependent on the level of activity (e.g., X gets upwards of 500 million posts per day) but also on the activity of users exchanging covert messages. All else being equal, the lower the base rate, the more difficult the task is for an adversary. This means that the higher the activity of covert messaging on the platform the worse it is for security, which may appear to be counter-intuitive but is simply a manifestation of base rate neglect~\cite{axelsson2000base}.
\begin{table}[!t]
    \caption{Expected outcomes on $100,000$ platform messages for covertexts generated with GPT-2 finetuned on \#news for varying base rates. FPR is set to $0.1\%$, which corresponds to a TPR of $2.7\%$. The last row is the posterior probability that a flagged post is in fact an covert message. \label{tbl:sec:impl} }
    %\vspace{-6pt}
    \centering
    \resizebox{.85\linewidth}{!}{\small %
    \begin{tabular}{l|l|l|l|}
    \cline{2-4}
    & \multicolumn{3}{c|}{Base rate} \\ \cline{2-4} 
                                                & $0.1\%$  & $0.01\%$ & $0\%$ \\ \hline \hline
    \multicolumn{1}{|l||}{Actual covert messages}      & 100     & 10      & 0       \\ \hline
    \multicolumn{1}{|l||}{Total messages flagged}     & 103      & 100       & 100      \\ \hline
    \multicolumn{1}{|l||}{False alarms}              & 100      & 100       & 100      \\ \hline
    \multicolumn{1}{|l||}{Missed detection}          & 97      & 10       & 0       \\ \hline
    \multicolumn{1}{|l||}{Covert messages flagged}     & 3      & 0       & 0       \\ \hline
    \multicolumn{1}{|l||}{Posterior probability}        & 0.0263 & 0.0027 & 0       \\ \hline
    \end{tabular}
    } % 
\end{table}

To evaluate this, we imagine deploying a system like MBFTE on a platform with a high background activity so that a realistic base rate may be $0.1\%$ or even $0.01\%$. We take results from~\cref{sec:security:naturalness} to have concrete numbers and consider an adversary who minimizes false positives at the cost of true positives (FPR=0.1\%, TPR = 2.7\%). What is the (posterior) probability that if the adversary flags a post as a potential covert message, the post is in fact a covert message? It depends on the base rate. What if the base rate is $0.1\%$? Bayes' theorem shows this probability to be $0.0263$, which is to say that for every $100$ messages flagged, we should only expect 2 or 3 of them to actually be covert messages (\cref{tbl:sec:impl}).

\begin{table}[!th]

\centering
\caption{Outcomes of user detection (1\% of platform users post covert messages). We assume each user posts $100$ messages from a single user account on the platform and simulate $10,000$ hypothetical platform users and their messages.}

\label{tbl:detecting-users-1perc}
\resizebox{.925\linewidth}{!}{\small %
\begin{tabular}{p{2.3cm}|l|l|l|l|l|l|}
\cline{2-6}
 & \multicolumn{5}{c|}{Base rate} \\ \cline{2-6} 
 & 0.01 & 0.05 & 0.1 & 0.2 & 0.4 \\ \hline \hline
TPR at FPR=0.1\% & 0.01 & 0.01 & 0.02 & 0.09 & 0.81 \\ \hline \hline
Detect. Threshold & 0.999 & 0.998 & 0.997 & 0.987 & 0.905 \\ \hline
True Positives & 26 & 69 & 78 & 94 & 99 \\ \hline
False Positives & 1908 & 3017 & 2090 & 918 & 146 \\ \hline
True Negatives & 7992 & 6883 & 7810 & 8982 & 9754 \\ \hline
False Negatives & 74 & 36 & 22 & 6 & 1 \\ \hline
Precision & 0.013 & 0.021 & 0.036 & 0.093 & 0.404 \\ \hline

\end{tabular}
} 
\end{table}

\subsection{Detecting Users}

What if the adversary seeks to detect users of a covert messaging system instead of covert messages? 
The adversary can leverage the fact that a user may post multiple covert messages using the same platform account. To analyze this, we observe that the relevant base rate is the proportion of a single user platform account's posts that are covert messages. On the one hand, this rate could be significantly higher than for detecting individual messages making detection easier. On the other hand, there are several ways for the covert messaging system to decrease this base rate. For example it can distribute the covert messages across multiple platform accounts, post messages generated with a different model, or post non-generated posts to these same accounts. 

We experimentally evaluate the success rate of an adversary who seeks to detect covert messaging user accounts on the platform. For this, we setup the experiment as follows. We consider a (hypothetical) set of $10,000$ platform user accounts, out of which only $1\%$ are sending covert messages. We assume that all platform users post $100$ messages and associate with each covert user a base rate that determines the proportion of covertexts among the $100$ messages posted by that platform user account (e.g, if the base rate is $0.1$ then only $10$ posts are covertexts). With this setup, the adversary's task is binary classification: predict whether a given set of $100$ messages comes from a covert user (i.e., whether the set contains at least one covertext).

We apply our strongest detector (RoBERTa model with fine-tuning known) independently on each message and then aggregate the predictions. We expect that if a user is posting covertexts the detector's predictions will have higher confidence values than for non-covert users. So we can combine the predictions by averaging the top-$q$ proportion of confidence values where $q$ is set to the base rate of covert users. The average is then compared to a threshold to make the final prediction. The detection threshold is set to maximize the difference between the true positive and false positive rates.

The results are shown in~\cref{tbl:detecting-users-1perc} where we determined the true/false positives/negative and precision values based on the detection threshold. Also, to capture an adversary who seeks to minimize false alarms we calculate the true positive rate for a false positive rate of $0.1\%$. We observe that for base rates lower than $0.2$ the true positive rate is quite low --- comparable to the true positive rate of the RoBERTa detector with known fine-tuning when detecting individual messages. More importantly, for low base rates the precision (i.e., proportion of predicted covert users that are actually posting covertexts) is low (e.g., only $3.6\%$ for a base rate of $0.2$). Most of the platform users that the adversary flags are false alarms. 

The task of detecting users of covert messaging is in some sense easier than detecting individual covert messages, but for relatively low base rates the rate of false alarms potentially renders this non-viable for adversaries. It is worth emphasizing that even with a ML-based detector with relatively high AUC, detecting individuals messages or covert message users is surprisingly unsuccessful.

\section{Limitations \& Future Directions}
The focus of our paper is on the challenges that arise when building a model-based covert messaging system from a steganographic construction. We explicate these challenges, discuss potential solutions, and surface various performance-security tradeoffs. Like all prior works, we do not address some deployment issues associated with rendezvous, key exchange, or link parameter agreement. 

In addition to the above omissions, this is a rapidly evolving field, and future work should re-examine several topics in their time frame. For instance, this work was done using GPT-2, but future work should consider switching to more advanced models such as GPT-3/4 to get more accurate comparisons against non-generated text. Additionally, MBFTE and other systems that use large-language models for covert messaging are currently unsuitable for low-end or legacy mobile devices, because those devices may lack the resources (e.g. memory) to support such models. 

Note that while our focus is embedding information into covertext using a generative model, there has been research on embedding information in other channels on public platforms~\cite{hu2021steganography,li2018lost,yang2020behavioral}. For instance, using the liking of certain posts, or posting behavior to communicate small amounts of information. While the capacity of such methods is small, many of the techniques and security considerations could be applied to future work.

There are other channels that adversaries could use to detect covert-messaging users that we have not discussed. An example is usage pattern analysis; the average user of the platform and a user of the covert-messaging system could have different posting frequencies or patterns. An adversary could examine post content, such as ensuring posts are coherent with each other, or fit the expected demographics of the poster. Alternatively, the adversary could use means other than looking at the platform messages (e.g., correlating IP addresses).

A system such as ours could be misused by bad actors. However, the same concerns apply to any other covert/anonymous communication systems such as Tor. Unfortunately, even good faith use may spread misinformation, as the generated posts may be taken as truth. Care should be taken to choose overt signals and finetunings that minimize the effect of potential misinformation. 

\section*{Acknowledgments}

We would like to thank researchers from Galois Inc., especially Dave Archer and Alex Malozemoff, for helpful feedback on an early version of this work. This work was supported in part by Galois, Inc. and an award under DARPA BAA \#HR001118S0052. The views, opinions, and/or findings expressed are those of the author(s) and should not be interpreted as representing the official views or policies of the Department of Defense or the U.S. Government.

\bibliography{references}{}
\bibliographystyle{plain}

%%%%%%%%%%%%%%%%%%%%%%%%%%%%%%%%%%%%%%%%%%%%%%%%%%%%%%%%%%%%%%%%%%%%%%%%%%%%%%%%
\end{document}